\documentclass[aps,prl,twocolumn, titlepage,showpacs]{revtex4}

\usepackage{graphicx}

\usepackage{dcolumn}

\usepackage{bm}
\usepackage{amsfonts}

\begin{document}

\title{Green-Kubo relation for viscosity tested using experimental data for a 2D dusty plasma}

\author{Yan Feng}
\email{yan-feng@uiowa.edu}
\affiliation{Department of Physics and
Astronomy, The University of Iowa, Iowa City, Iowa 52242, USA}
\author{J. Goree}
\affiliation{Department of Physics and Astronomy, The University
of Iowa, Iowa City, Iowa 52242, USA}
\author{Bin Liu}
\affiliation{Department of Physics and Astronomy, The University
of Iowa, Iowa City, Iowa 52242, USA}
\author{E.~G.~D.~Cohen}
\affiliation{The Rockefeller University, 1230 York Avenue, New
York, New York 10065, USA} \affiliation{Department of Physics and
Astronomy, The University of Iowa, Iowa City, Iowa 52242, USA}

\date{\today}

\begin{abstract}

The theoretical Green-Kubo relation for viscosity is tested using
experimentally obtained data. In a dusty plasma experiment,
micron-size dust particles are introduced into a partially-ionized
argon plasma, where they become negatively charged. They are
electrically levitated to form a single-layer Wigner crystal,
which is subsequently melted using laser heating. In the liquid
phase, these dust particles experience interparticle electric
repulsion, laser heating, and friction from the ambient neutral
argon gas, and they can be considered to be in a nonequilibrium
steady state. Direct measurements of the positions and velocities
of individual dust particles are then used to obtain a time series
for an off-diagonal element of the stress tensor and its time
autocorrelation function. This calculation also requires the
interparticle potential, which was not measured experimentally,
but was obtained using a Debye-H\"{u}ckel-type model with
experimentally determined parameters. Integrating the
autocorrelation function over time yields the viscosity for
shearing motion amongst dust particles. The viscosity so obtained
is found to agree with results from a previous experiment using a
hydrodynamical Navier-Stokes equation. This comparison serves as a
test of the Green-Kubo relation for viscosity. Our result is also
compared to the predictions of several simulations.

\end{abstract}

\pacs{52.27.Lw, 52.27.Gr, 66.20.-d, 83.85.Jn, 05.60.Cd}\narrowtext

\maketitle

\section {I.~INTRODUCTION}

There are various two-dimensional (2D) physical systems that allow
direct observation of individual particle dynamics. These systems
include colloidal suspensions~\cite{Murray:90}, granular
materials~\cite{Reis:06}, electrons on a liquid helium surface in
a Wigner lattice~\cite{Grimes:79}, ions confined magnetically in a
Penning trap~\cite{Mitchell:99}, and single-layer dusty
plasmas~\cite{Shukla:02, Morfill:09, Piel:10, Feng:10}. In these
systems, the relevant particles collide with their neighbors
frequently, so that  momentum and energy are transported from one
place to another. (In all these 2D physical systems, motion is not
purely 2D, but usually includes some limited out-of-plane motion,
so that the systems are often described as quasi-2D.)

Shear viscosity, $\eta$, is a transport coefficient that
characterizes the momentum flux perpendicular to a velocity
gradient. Sustaining the velocity gradient requires the
application of a shear stress, which corresponds to an
off-diagonal element of a stress tensor. The hydrodynamical
definition of shear viscosity is the ratio of this off-diagonal
element and the velocity gradient~\cite{Batchelor}. As a measure
of dissipation, viscosity is useful, for example, in describing
the damping of shear waves~\cite{Murillo:00, Salin:07, Kaw:98}.

The Green-Kubo relation for viscosity, as described in Sec.~II,
allows a calculation of viscosity using as its input a time series
record of the motion of individual particles~\cite{G, K,
Hansen:86}. The relation is based on fluctuations, not a
macroscopic velocity gradient. The Green-Kubo relation has been
widely used in molecular dynamics (MD) {\it simulations}, for
example~\cite{Donko:07}. However, there is a need for a test of
the Green-Kubo relation for viscosity using an input of
experimental data. In our search of the literature we have not
found any such test, nor have we found any previous use of the
Green-Kubo relation with an input of experimental data to
determine viscosity.

Viscosity is most commonly determined in experiments using  a
macroscopic velocity gradient. For example, in a
rheometer~\cite{Pal:00}, a stress is applied by a moving boundary,
causing the liquid to flow with a velocity gradient, so that the
viscosity can be determined using its hydrodynamical definition.
Another experimental approach, which is used with experimental
data from colloidal suspensions and other soft materials without
macroscopic velocity gradients, is the measurement of the
mean-square displacement (MSD) of individual particles and the
assumption that the Stokes-Einstein relation is valid, allowing a
calculation of the viscosity from the measured MSD~\cite{Mason:95,
Crocker:00}, as discussed in Sec.~II. All of these experimental
methods are different from the use of the Green-Kubo relation.

It has been questioned whether transport coefficients exist in 2D
systems. Molecular dynamical simulations suggested that the
diffusion, viscosity, and thermal conductivity coefficients would
not exist in a 2D system of {\it{hard disks}}~\cite{Alder:67}.
This result led to theoretical investigations that indicated the
time integral in the Green-Kubo relations diverges and the
corresponding transport coefficients in 2D systems are
non-existing~\cite{Ernst:70}. We will discuss this issue of
convergence in Sec.~IV~E.

In this paper, we test the Green-Kubo relation with an input of
experimental data from a quasi-2D dusty plasma. The dust particles
in a dusty plasma, unlike hard disks, have a relatively long-range
interparticle interaction.

Our results and conclusions in this paper are intended to have a
usefulness that extends beyond the area of dusty plasmas.
Accordingly, we have attempted to make our presentation accessible
to scientists who are not specialists in that area. The
experimental data used in this paper are from an experiment by
Feng {\it et al.}~\cite{Feng:10}. We will introduce the concepts
of dusty plasmas, along with some key aspects of the experiment,
below as well as in Sec.~III. Further details of the experiment
can be found in~\cite{Feng:10}.

A dusty plasma is a {\it four-component} mixture consisting of
micron-size particles of solid matter, a gas of neutral atoms,
free electrons, and free positive ions. We will refer in this
paper to the particles of solid matter as ``dust particles.'' In
experiments, the gas is often argon, and so are the ions. The
electrons and positive ions are present because of electric
fields, provided by a power supply, which lead to a weak
ionization of the neutral argon gas. Because of the much larger
mobility of the electrons, as compared to the ions, many more
electrons are collected on the dust particles, so that a dust
particle develops a large negative electric charge, comparable to
thousands of elementary charges~\cite{Goertz:1989}. The electrons
and ions that surround a negatively-charged dust particle are
disturbed, resulting in a screening layer with a surplus of
positive ions that surrounds the dust particle. Because of the
complicated interactions amongst its four components, a dusty
plasma is sometimes termed a ``complex plasma''
\cite{Samsonov:2000}. While the gas, electrons and ions fill a 3D
vacuum chamber, the dust particles do not. They are levitated
against the downward force of gravity by a strong vertical
electric field. In the experiment~\cite{Feng:10}, enough dust
particles were introduced to fill a single horizontal layer, but
not enough to begin filling a second layer. The dust particles
were not in contact with any solid boundary, but they suffer a
friction due to the surrounding neutral argon gas. In this paper,
we will always consider the gas atoms as a whole to be a
continuum, but we will consider the dust particles as individual
entities.

The  experiment~\cite{Feng:10} can be described as quasi-2D. The
dust particles, although they are three dimensional, are arranged
in two dimensions. It was verified, using video observations, that
the dust particle motion in the vertical direction was extremely
limited, as compared to that in the horizontal direction, so that
dust particles moved past one another only as a result of their
horizontal motion.

The neutral argon gas is rarefied in the
experiment~\cite{Feng:10}. At a density five orders of magnitude
less than in a standard atmosphere, collisions amongst gas atoms
have a long mean free path, on the order of 1 cm. The effect of
those collisions is diminished even more because the dust
particles occupy only a thin 2D layer, so that a gas atom that
collides with a dust particle is likely to be knocked out of the
layer of the dust particles~\cite{Feng:11}. Thus, no significant
transfer of momentum between two dust particles can occur due to
the first one colliding with a gas atom which then collides with
the second one. The only interaction between dust particles and
gas that we must consider is the frictional drag force $F_f$ on
the dust particle, which is proportional to the relative velocity
between the dust particle and the gas as a whole.
In~\cite{Feng:10} the gas flow was negligibly slow.

Since it is only the motion of the dust particles that will be of
interest here, we will simplify our description of the
four-component mixture. The dust particles are assumed to interact
amongst themselves with a screened Coulomb repulsion, as discussed
in Sec.~IV. The role of the electrons and ions is then only to
modify the interparticle potential and provide the screening.
Thus, in our simplified description of the four-component mixture
we consider only a {\it binary mixture}: first, moving charged
dust particles whose interaction potential is a screened Coulomb
repulsion, and second a stationary neutral gas that exerts a
frictional drag on moving dust particles.

This reduction of a four-component mixture to a binary mixture, in
which all the properties of the electrons and ions are contained
in the screening, has been used previously in theoretical
descriptions of dusty plasmas, for example in the analysis of wave
propagation~\cite{Wang:01}. When using this binary-mixture
description, one could consider a charged dust particle as a
``dressed particle''~\cite{Joyce:01} consisting of a micron-size
solid core that is negatively charged and a larger surrounding
screening region that is positively charged. The center of this
dressed particle corresponds to what is observed experimentally by
video microscopy.

In the experiment~\cite{Feng:10}, the repulsion between dust
particles was so strong that the dust particles self-organized in
a solid-like arrangement called a Wigner
crystal~\cite{Bonsall:77}. In order to study a liquid and its
viscosity, this solid was melted by increasing the kinetic energy
of the dust particles, which was done by using the laser-heating
method~\cite{Nosenko:06}.

When we refer in this paper to viscosity, it is only the motion of
the dust particles that we directly take into account. In our
simplified description of a dusty plasma, treating it as a binary
mixture of dust particles and gas, we do not consider the momentum
carried by electrons and ions. Moreover, a transfer of momentum
between two dust particles does not occur due to the first dust
particle colliding with a gas atom which then collides with the
second, as discussed above. Thus, in our simplified binary-mixture
description, the viscosity describes motion of only dust
particles~\cite{Einstein:11}.

Previously, viscosity was studied in other dusty plasma
experiments by applying a macroscopic shear stress using laser
manipulation~\cite{Juan:01, Nosenko:04, Gavrikov:05, Vorona:07} to
generate a macroscopic velocity gradient, and using a
hydrodynamical approach to calculate the viscosity based on the
measured velocity profile of the dust particles. In the test
reported in this paper, we will compare the hydrodynamical result
of~\cite{Nosenko:04} to the viscosity determined using a
theoretical Green-Kubo relation with an input of data from the
experiment of~\cite{Feng:10}, which was performed without a
macroscopic velocity gradient.

In Section~II, the Green-Kubo relation for viscosity is reviewed.
In Sec.~III, we provide further details of the
experiment~\cite{Feng:10}. In Sec.~IV, we introduce how we use the
Green-Kubo relation with an input of experimental data. In Sec.~V,
we report our MD simulations of the experiment~\cite{Feng:10}. We
determine the viscosity in Sec.~VI, and in Sec.~VII this result
using the Green-Kubo relation is compared to the results of a
previous experiment using a hydrodynamical
method~\cite{Nosenko:04}. In Sec.~VII we also provide a comparison
to the results of previous computer simulations~\cite{Donko:06,
Feng:11, Liu:05}.

\section {II.~GREEN-KUBO RELATION}

To obtain transport coefficients such as diffusion, viscosity, and
thermal conductivity in a liquid, Green-Kubo relations~\cite{G, K,
Hansen:86, Donko:09, Liu:05} are often used. Their required inputs
include time series for the positions and velocities of particles.
The simplest of these Green-Kubo relations is the one for
diffusion. It can be derived easily using the physical assumption
that the MSD for fluctuating particle displacements is
proportional to the diffusion coefficient and the
time~\cite{Hansen:86}. The derivation of the Green-Kubo relation
for viscosity is less trivial, and it is based on the fluctuating
stress, not an MSD~\cite{McQuarrie:76}. Here we review the
standard Green-Kubo relation for calculating viscosity, in three
steps, as it is used for all kinds of liquids, not just dusty
plasmas in a liquid phase.

In the first step, an off-diagonal element of the stress tensor
$P_{xy}(t)$ is defined by
\begin{equation}\label{SS}
{P_{xy}(t)=
\sum_{i=1}^N\left[mv_{ix}v_{iy}-\frac{1}{2}\sum_{j\not=i}^N\frac{x_{ij}y_{ij}}{r_{ij}}\frac{\partial
\Phi(r_{ij})}{\partial r_{ij}}\right],}
\end{equation}
where $i$ and $j$ denote different particles, $N$ is the total
number of  particles of mass $m$, $\mathbf{r}_{i} = (x_i,y_i)$ is
the position of particle $i$, $x_{ij}=x_i-x_j$, $y_{ij}=y_i-y_j$,
$r_{ij}=|\mathbf{r}_i-\mathbf{r}_j|$, and $\Phi(r_{ij})$ is the
interparticle potential energy. Although not indicated in
Eq.~(\ref{SS}), the positions and velocities of particles vary
with time, and this accounts for the time dependence of
$P_{xy}(t)$. In the second step, an autocorrelation function of $
P_{xy}(t)$ is calculated as
\begin{equation}\label{SACF}
{C_{\eta}(t)= \langle P_{xy}(t)P_{xy}(0) \rangle.}
\end{equation}
We will refer to $C_{\eta}(t)$ as the stress autocorrelation
function (SACF). The brackets $\langle \cdot\cdot\cdot \rangle$
denote an average over an equilibrium ensemble, which in practice
is often replaced by an average over different initial conditions.
In the third step, the SACF is integrated over time to yield the
viscosity $\eta$; for a 2D system the result is
\begin{equation}\label{eta}
{\eta=\frac{1}{A k_B T}\int^\infty_0 C_{\eta}(t)dt,}
\end{equation}
where $A$ is the area of the 2D system and $T$ is its temperature.
Equation (\ref{eta}) combined with Eq.~(\ref{SACF}) represent the
Green-Kubo relation for viscosity in 2D. Similar Green-Kubo
relations can be written for diffusion and thermal
conductivity~\cite{Hansen:86, Donko:09}.

Viscosity $\eta$ and mass density $\rho$ have different dimensions
in 2D and 3D. The units of $\eta$ are $\rm{kg \, s^{-1}}$ in 2D,
and $\rm{kg \, m^{-1}s^{-1}}$ in 3D. Correspondingly, in the
denominator of Eq.~(\ref{eta}) we have replaced the usual volume
for a 3D system with an area $A$ for the 2D system. In 2D, the
areal mass density is $\rho=nm$ with units of ${\rm kg \,
m^{-2}}$, where $n$ is the areal number density. We will report
results for the kinematic viscosity $\eta/\rho$, which has the
same units in 2D and 3D.

While Green-Kubo relations have been commonly used in computer
simulations, their use with experimental data is uncommon. We are
aware of only one previous calculation of any transport
coefficient using the input of experimental data in a Green-Kubo
relation. Using data from a dusty plasma experiment, Vaulina {\it
et al.}~\cite{Vaulina:08, Vaulina:08_2} obtained the {\it
diffusion} coefficient using its Green-Kubo relation, which is a
time integration of the velocity autocorrelation function.
Calculating this autocorrelation function required an input of
dust particle velocities, which were determined from experimental
measurements of dust particle positions. In principle, the
approach of Vaulina {\it et al.} of using a Green-Kubo relation to
obtain the diffusion coefficient could be extended to other
transport coefficients: the viscosity, thermal conductivity, and
bulk viscosity.

In this paper, we use the the Green-Kubo relation for {\it
viscosity} with an input of experimental data. A comparison of our
result for viscosity with values determined in a previous
experiment using a hydrodynamic method with a velocity
gradient~\cite{Nosenko:04} will serve as an experimental test of
the Green-Kubo relation for viscosity.

Besides the Green-Kubo relation, microrheology is another method
to obtain the viscosity of fluids without a macroscopic velocity
gradient~\cite{Mason:95, Crocker:00}. In this approach, the MSD of
individual microparticles is measured, and the Stokes-Einstein
relation is assumed to be valid~\cite{Mason:95, Crocker:00}. The
Stokes-Einstein relation~\cite{Jones} is a combination of the
Stokes law, which is a hydrodynamic model for viscous flow at a
low Reynolds number, and the Einstein relation, which relates a
diffusion coefficient for Brownian motion and a frictional force.
This MSD-based method has been used in physical systems like
colloidal suspensions, where a microparticles's motion is
overdamped due to the surrounding liquid solvent~\cite{Mason:95,
Crocker:00}.

The Green-Kubo relation for viscosity is different from MSD-based
approaches of determining viscosity. The derivation of the
Green-Kubo relation centers on the fluctuations of the stress
$P_{xy}$, and it does not rely on the validity of a diffusion
coefficient. The Stokes-Einstein relation, while having great
utility for many physical systems, is known to fail for others
such as supercooled liquids~\cite{Rossler:90}. Moreover, it is
possible that a physical system can have a valid viscosity
coefficient but lack a valid diffusion coefficient, for example
due to superdiffusion, as has been suggested for 2D systems, such
as Yukawa liquids~\cite{Liu:05, Donko:09}.

\section {III.~EXPERIMENTAL INPUT}

Before reviewing the experiment~\cite{Feng:10} we will discuss a
few properties of dusty plasmas and their relevant length and time
scales.

When dust particles have a charge of several thousand elementary
charges, their interparticle potential energy $\Phi$ can be larger
than their kinetic energy. In this case, the collection of dust
particles is said to be a ``strongly-coupled
plasma''~\cite{Ichimaru, Kalman:04}. A measure of strong coupling
is the Coulomb coupling parameter $\Gamma \equiv
(Q^2/4\pi\epsilon_0a k_B T)$, where $Q$ is the charge of the dust
particle, $\epsilon_0$ is the permittivity of free space, $a$ is a
typical interparticle distance as defined below, and $T$ is the
kinetic temperature of the dust particles. The Coulomb coupling
parameter is essentially a ratio of interparticle potential energy
and kinetic energy. A plasma is strongly coupled when $\Gamma >
1$, and it can behave like a liquid or a Wigner crystal. In a
dusty plasma, the dust particles are usually strongly coupled due
to their large charge $Q$. The electrons and the ions have a much
smaller charge, and in most plasmas they are not strongly coupled,
unless great efforts are made to cool them to low
temperatures~\cite{Mitchell:99}.

The {\it length scales} of a dusty plasma include the screening
length $\lambda_D$ and the typical distance between dust
particles.
In the dusty plasma literature,
the typical distance between dust particles is commonly reported
either as the lattice constant $b$ for a Wigner crystal, or as
$a=(n\pi)^{-1/2}$ for a liquid, where $n$ is the areal number
density. In the literature for strongly-coupled plasmas, $a$ is
called the 2D Wigner-Seitz radius~\cite{Kalman:04}. All three of
these length scales, $\lambda_D$, $b$ and $a$, are typically on
the order of 1~mm. The diameter of a dust particle is typically a
few microns, which is much smaller than any of these length
scales, and also much smaller than the mean free path for
collisions amongst the rarefied argon gas atoms.

The {\it time scales} of a dusty plasma include measures of
collective motion amongst the dust particles, and of the
frictional drag experienced by a dust particle due to the gas as a
whole. The former is quantified by the 2D plasma frequency for
dust particles, $\omega_{pd}=
(Q^2/2\pi\epsilon_0ma^3)^{1/2}$~\cite{Kalman:04}, where the
subscripts $p$ and $d$ refer to plasma and dust, respectively. The
frictional drag experienced by a dust particle due to the neutral
argon gas is quantified by a friction coefficient, which is
commonly defined in the literature for dusty plasmas as $\nu_f =
F_f / mv $, where $F_f$ is the gas friction force experienced by
one dust particle and $mv$ is the momentum of the same particle.
This coefficient has the dimension of inverse time, and a typical
value for a micron-size dust particle in a rarefied gas is $\nu_f
\approx 1~\rm{s^{-1}}$. If the gas is sufficiently rarefied,
$\omega_{pd}
> \nu_f$, so that the dust particle motion is said to be underdamped.

We now summarize some aspects of the experiment~\cite{Feng:10}
that are relevant for our analysis. The dust particles were
polymer microspheres of $8.1~{\rm \mu m}$ diameter. They were
levitated by a vertical dc electric field to form a single
horizontal layer, as sketched in Fig.~1. Radial dc electric fields
provided horizontal confinement so that the dust particles filled
a circular region of diameter 52~mm.

For each experimental run, one movie of dust particle motion was
recorded. A Phantom v5.2 high-speed camera viewed the dust
particles from above. It was operated at 250 frames/s, so that
data were recorded with a time interval $\Delta t = 4~{\rm{ms}}$.
The duration of a movie, $20~{\rm s}$, was limited by the camera's
memory. The camera's field-of-view (FOV) included $N \approx~2100$
dust particles. Two of the required inputs for Eq.~(\ref{SS}), the
positions and velocities of the dust particles, were obtained
using the moment method~\cite{Feng:07} and particle
tracking~\cite{Feng:11RSI}. The interparticle distance, averaged
over the camera's FOV, was characterized by $b=0.67~{\rm mm}$ in
the Wigner crystal, and $a=0.35~{\rm mm}$ in the liquid.

A total of seven runs were performed. Three runs without laser
heating were made to determine the interparticle potential energy
in the Wigner crystal. Four runs were done with laser heating to
make the liquid, and they provide the data that we will use here
as the input for the Green-Kubo relation, to determine viscosity
in the liquid phase.

The laser heating method~\cite{Wolter:05} uses the
radiation-pressure due to laser beams, which are directed toward
the dust particles by scanning mirrors~\cite{Melzer:2000}, which
we configured as in~\cite{Nosenko:06, Liu:08}. Laser heating
increases the kinetic temperature of dust particles without
changing their charge. The velocity distribution function has been
observed to be nearly Maxwellian~\cite{Nosenko:06}. In the
experiment~\cite{Feng:10}, the kinetic temperature $T$, calculated
from the mean-square velocity~\cite{Feng:08}, was $2.5\times
10^4~K$ in the liquid (with laser heating), and $10^3~K$ in the
Wigner crystal (without laser heating). The temperature was nearly
uniform spatially, which is desirable because viscosity varies
with temperature.

\section {IV.~USING THE GREEN-KUBO RELATION WITH EXPERIMENTAL INPUT DATA}

Calculations of viscosity with the Green-Kubo relation
Eqs.~(\ref{SS})-(\ref{eta}) must be adapted  to use the
experimental data as input, due to four difficulties. First, the
experiment~\cite{Feng:10} provided direct measurements of $x_i$,
$y_i$, $v_{ix}$, and $v_{iy}$, but not of the interparticle
potential energy $\Phi(r_{ij})$. Second, the camera has a finite
FOV, so that we do not have data for all dust particles in the 2D
layer. Third, the motion of dust particles may include a local
macroscopic flow, i.e., a non-zero time-average velocity. Fourth,
the data for the positions and velocities of the dust particles
are recorded as a time series of a finite duration, so that the
integral in Eq.~(\ref{eta}) must have a finite limit. We will take
all this into account by making a number of approximations in the
Green-Kubo calculations. We will next describe these
approximations as well as discuss the validity of the results that
are obtained.

\subsection {A. Interparticle potential}

To solve the first difficulty, a lack of experimental measurements
of the interparticle potential energy $\Phi(r_{ij})$, we will use
a model for these energies when calculating the off-diagonal
element of the stress tensor $P_{xy}$. For a single-layer dusty
plasma like ours, models that have been tested successfully
include an isotropic repulsion according to the 3D
Debye-H\"{u}ckel potential~\cite{Konopka:00}
\begin{equation}\label{DH}
{\phi(r_{ij}) = \frac {Q}{4\pi\epsilon_0} \frac{{\rm
exp}(-r_{ij}/\lambda_D)}{r_{ij}},}
\end{equation}
as well as more complicated isotropic~\cite{Vaulina:2010} and
non-isotropic interactions~\cite{Schweigert:98-Lampe:00}. Here we
will use the Debye-H\"{u}ckel potential, Eq.~(\ref{DH}), where $i$
and $j$ are dust particles of charge $Q$ separated by a distance
$r_{ij}$, and $\lambda_D$ is the screening length due to electrons
and ions. The corresponding potential energy is $\Phi(r_{ij}) = Q
\phi(r_{ij})$. In the literature for dusty plasmas, it is common
to name Eq.~(\ref{DH}) after Yukawa rather than Debye and
H\"{u}ckel. This potential has been used in theoretical and
simulation studies of viscosity in strongly coupled plasmas, for
example~\cite{Donko:09, Murillo:00_2, Saigo:02, Salin:03,
Faussurier:04}.

Two parameters in Eq.~(\ref{DH}), $Q$ and $\lambda_D$, were
determined in the experiment~\cite{Feng:10} using a phonon-spectra
method~\cite{Nunomura:02, Nunomura:05} in the Wigner crystal. For
the three experimental runs without laser heating, position and
velocity measurements were used to compute the phonon spectra,
which were compared to theoretical wave dispersion
relations~\cite{Wang:01} that assume Eq.~(\ref{DH}), yielding
$\lambda_D=0.70\pm0.14~{\rm mm}$ and $Q/e=-6000\pm600$, where $e$
is the elementary charge. Using these two values along with
measured particle positions and velocities, we can calculate
$P_{xy}$ and then the SACF~\cite{error-test}.

Having determined $Q$ and $\lambda_D$, we can calculate the values
of other parameters. We find that the Coulomb coupling parameter
is $\Gamma =68$ for the liquid of dust particles. The
dimensionless particle spacing~\cite{Kalman:04} is $\kappa \equiv
a/\lambda_D=0.5\pm0.1$. We also calculated
$\omega_{pd}=30\pm3~{\rm s^{-1}}$. We note that $\omega_{pd}$ is
significantly larger than the gas friction coefficient
$\nu_f=2.4~{\rm s^{-1}}$, indicating that dust particle motion is
underdamped.

\subsection {B.~Finite field of view}

To solve our second difficulty, the finite field of view (FOV) of
the camera that limits us to recording data for only a portion of
the dust particles, we will {\it cut off} the interparticle
potential in Eq.~(\ref{DH}) at large distances. In addition, we
will {\it divide} the FOV into inner and outer regions, Fig.~2.
The cutoff is done at a large interparticle distance of $5b$,
where the exponential in Eq.~(\ref{DH}) is $<10^{-2}$.

The FOV is divided into inner and outer regions because the
potential energy of a dust particle cannot be obtained
meaningfully if it is located near the edge of the FOV, due to
interactions with dust particles of unknown positions outside the
FOV. Therefore, dust particles in the outer region are used only
to calculate the potential energies $\Phi$ of dust particles in
the inner region. In other words, when calculating $P_{xy}$, we
limit the dust particles $i$ to those that are located within the
inner region, and we account for their interaction with other dust
particles $j$ located in {\it both} the inner and outer regions,
as shown in Fig.~2. The outer region has a width of $5b$ to allow
for a cutoff radius of $5b$, and the inner region is $22.0b \times
23.7b$ (i.e., $14.8\times 15.9~{\rm mm^2}$). Because of this
different treatment of dust particles in the inner and outer
regions, we calculate $P_{xy}$ as
\begin{equation}\label{SSexp}
{P_{xy}=
\sum_{i=1}^M\left[mv_{ix}v_{iy}-\frac{1}{2}\sum_{j\not=i}^N\frac{x_{ij}y_{ij}}{r_{ij}}\frac{\partial
\Phi(r_{ij})}{\partial r_{ij}}\right],}
\end{equation}
where the stress $P_{xy}$ is (implicitly) a function of time. All
time series data are recorded at a time interval $\Delta t$ =
4~ms. For the experiment~\cite{Feng:10}, $M \approx 600$ dust
particles are in the inner region, while $N \approx 2100$ are in
both regions combined. Both $N$ and $M$ fluctuate slightly with
time, as dust particles move across the edges of the regions, but
not enough to affect the result for viscosity significantly.

\subsection {C.~Non-zero time averages}

Our third difficulty to solve is that the average velocity is not
zero, due to finite macroscopic flow velocities of the dust
particles, despite efforts that were made in the
experiment~\cite{Feng:10} to avoid them. When computing the
stress, the dust particle velocities are assumed to fluctuate
about an average value of zero. In fact, a non-zero average
velocity would contribute an unwanted constant value to the
$P_{xy}$ time series, when computed using Eq.~(\ref{SS}) or
(\ref{SSexp}), which would cause the SACF, $C_{\eta}$, to decay to
a non-zero
 value and introduce an unphysical contribution
to the calculated viscosity. To solve this difficulty, without any
approximation, we subtract from ${P}_{xy}(t)$ its time-average
value, yielding then the fluctuating portion $\tilde{P}_{xy}(t)$.
We then replace Eq.~(\ref{SACF}) for the SACF by
\begin{equation}\label{SACFEXP}
{C_{\eta}(t)= \langle \tilde{P}_{xy}(t+t_0)\tilde{P}_{xy}(t_0)
\rangle_0.}
\end{equation}
Here, the brackets $\langle \cdot\cdot\cdot \rangle_0$ in
Eq.~(\ref{SACFEXP}) denote an average over various initial times
$t_0$, if the data are from a single run.

Our results for the SACF, for the four runs with laser heating,
are shown in Fig.~3. All four runs show the same general trends,
which resemble those seen in MD simulations as in~\cite{Liu:05}.

\subsection {D.~Integration limit}

To solve the fourth difficulty, the finite time duration of data,
the viscosity $\eta$ is computed with a finite upper limit in the
time integral of the SACF. In principle the upper limit should be
infinite, as in Eq.~(\ref{eta}), but we
 use
\begin{equation}\label{etaEXP}
{\eta=\frac{1}{A k_B T}\int^{t_I}_0 C_{\eta}(t)dt,}
\end{equation}
and we follow the practice used in MD simulations of choosing the
integration limit $t_I$ as the time when $C_{\eta}(t)$ crosses
zero~\cite{Donko:10}, as shown in Fig.~3. Because our time series
$C_{\eta}(t)$ is noisy, we count a zero crossing only if it
results in $C_{\eta}(t)$ remaining negative for at least $2 \Delta
t$. In Eq.~(\ref{etaEXP}), $A$ is the area of the inner region.

\subsection {E.~Validity}

For using the Green-Kubo relation to calculate viscosity with
experimental input data, we should ask whether the approach is
valid, when it is used for a dusty plasma. We will mention three
questions.

First, we note that the Green-Kubo relations are strictly speaking
for the thermodynamic limit, where the number of dust particles
and the system size tend to infinity while keeping the number
density constant. In fact, the experiment has only thousands of
dust particles. However, we believe that our experimental system
size is large enough to use a Green-Kubo relation, as indicated by
our system-size tests in Sec.~V.

Second, we must consider the distinction between equilibrium and
nonequilibrium systems.  While our laser-heated dusty plasma is in
a steady state, it is not in equilibrium. The collection of dust
particles is best described as a driven-dissipative
system~\cite{Liu:08}, where the driving is mainly provided by the
laser beams, and the dissipation is provided by gas-dust
collisions as well as the dust-particle viscosity. Despite these
nonequilibrium conditions, however, the velocity distribution
function for the dust particles has been observed to be nearly
Maxwellian~\cite{Nosenko:06}. Thus, the statistics are close to
those of an equilibrium system, which motivates us to use the
Green-Kubo relation.

Third, we must ask whether long-time tails in the correlation
function prevent the convergence of its integral, as was predicted
theoretically~\cite{Ernst:70} for 2D systems with hard-disk
interparticle interactions. More recently, for a 2D system with a
Debye-H\"{u}ckel potential, simulations by Donk\'{o} {\it et
al.}~\cite{Donko:09} indicated that long-time tails occur in some
but not all cases. In particular, they reported that the SACF
decays fast enough that its time integral $\eta$ converges for a
liquid at temperatures near the melting point, but not at absolute
temperatures far above the melting point.

\section {V. Simulations}

To assess three sources of error mentioned in Sec. IV, {\it viz.}
the potential cutoff, the FOV division, and the finite system
size, we used MD simulations based on the Langevin
equation~\cite{Reif:1965}. In these simulations we numerically
integrate the Langevin equation to obtain the motion of each dust
particle. This equation gives the force acting on a dust particle
as three terms: a sum of an electric force due to all other dust
particles using Eq.~(\ref{DH}), a mean friction $F_f$ due to the
gas as a whole, as well as Gaussian random forces around this
mean, to model the collisions of the dust particle with the
neutral gas atoms~\cite{Gunsteren:1982, Ott:2009}.

Dust particle positions, velocities, and interaction energies were
recorded at each time step of $0.019~\omega_{pd}^{-1}$. We used $N
= 4096$ particles in a 2D rectangular box with periodic boundary
conditions. In a 2D Langevin MD simulation with a Debye-H\"{u}ckel
potential like ours, the equations of motion have three
dimensionless parameters: the friction coefficient
$\nu_f/\omega_{pd}$, the Coulomb coupling parameter $\Gamma$, and
the dimensionless particle spacing $\kappa$. To mimic the
experiment~\cite{Feng:10}, we used $\nu_f / \omega_{pd}=0.08$,
$\Gamma=68$, and $\kappa=0.5$.

To test the effect of the cutoff, we carried out simulations at
two cutoff distances to estimate the systematic error introduced
in the last term of Eq.~(\ref{SSexp}). We found that for a cutoff
of $5 b$, which we use in this paper, the viscosity result $\eta$
was reduced by less than $5\%$, as compared to the result for a
much larger cutoff of $13 b$.

To test for errors arising from the division of the FOV, we
compared results with and without the division, using
Eq.~(\ref{SSexp}) and Eq.~(\ref{SS}), respectively. We found the
viscosity differed only negligibly.

Finally, to test for the effects of finite system size, we
compared our Langevin MD simulation results for two sizes. Results
for the larger system size of $4096$ particles are reported
in~\cite{Feng:11}. For the smaller size, to mimic the
experiment~\cite{Feng:10}, we used 48 simulation runs with $M =
600$ particles, as in the inner region for the
experiment~\cite{Feng:10}, for a
duration~\cite{footnote-memory-time} of $607~\omega_{pd}^{-1}$.
Comparing these two results for $\eta$, we find no
statistically-significant system-size effects. This test, shown in
Table~I, gives us confidence that the number of dust particles in
~\cite{Feng:10} was not so small as to preclude using the
Green-Kubo relation.

\section {VI.~RESULT FOR VISCOSITY USING THE GREEN-KUBO RELATION WITH EXPERIMENTAL
DATA}

We now present our result for the viscosity $\eta$ using the
Green-Kubo relation, Eqs.~(\ref{SSexp})-(\ref{etaEXP}). We report
the kinematic viscosity, $\eta/\rho$, to allow convenient
comparison to other experiments and simulations.

We find $\eta/\rho = 0.16 \pm 0.02$, in units of $a^2\omega_{pd}$.
The value of the normalization factor for the
experiment~\cite{Feng:10} is $a^2\omega_{pd} = 3.7 \times
10^{-6}~{\rm m^2/s}$, while the areal mass density is $\rho = 1.1
\times 10^{-6}~{\rm kg/m^2}$. The value of $0.16$  is the mean of
the four experimental runs in the presence of laser heating, as
plotted in Fig.~3. The error estimate of $\pm~0.02$, calculated as
the standard deviation of the mean, indicates  the run-to-run
random variation.

\section {VII. Discussion}

\subsection{A. Test of the Green-Kubo relation}

Our most significant result is a test of the Green-Kubo relation
for viscosity, using an input of experimental data. We perform
this test by comparing our result $\eta/\rho = 0.16 \pm~0.02$ to
the previously-reported result, from a dusty plasma
experiment~\cite{Nosenko:04} that used a hydrodynamical approach.

In the experiment of Nosenko and Goree~\cite{Nosenko:04}, the dust
particles flowed in their quasi-2D layer with a macroscopic
velocity gradient, allowing a determination of viscosity using a
Navier Stokes equation of motion for the local flow velocity of
the dust particles.
Differently from~\cite{Nosenko:04}, in our analysis here we
consider the dust particles only as individual particles, while
in~\cite{Nosenko:04} data for individual particles were averaged
to allow considering the dust particles as a continuum, as is
necessary for a hydrodynamical approach.
Both experiments,~\cite{Feng:10} and \cite{Nosenko:04}, were
performed in the same chamber, and both had a quasi-2D layer of
dust particles. The reported values of $\kappa$ were nearly the
same, $\kappa = 0.53$ for~\cite{Nosenko:04} and 0.5
for~\cite{Feng:10}. The electrical interactions among dust
particles was much stronger than gas friction, $\omega_{pd} \gg
\nu_f $, in both experiments. A difference in the experimental
conditions was that the two laser beams were manipulated
differently so that in~\cite{Nosenko:04} they produced a
macroscopic velocity gradient, while in~\cite{Feng:10} there was
heating without a macroscopic velocity gradient.

Considering the complexity of the dusty plasma and our
simplifications in describing it, we cannot expect  our result to
agree with the results of~\cite{Nosenko:04} better than within
about a factor of two. In fact, we find agreement within this
factor when comparing our result $\eta/\rho = 0.16 \pm~0.02$ using
the Green-Kubo relation and the range of values reported by
Nosenko and Goree~\cite{Nosenko:04}, Fig.~4.

\subsection{B. Comparison to simulations}

We also compare our result from the Green-Kubo relation to the
available data for viscosity from the simulation literature,
Fig.~5.  All these data are from 2D MD simulations with a
Debye-H\"{u}ckel potential, and most of them use the Green-Kubo
relation, except the nonequilibrium simulation of~\cite{Donko:06},
which used a so-called nonequilibrium simulation method to produce
a macroscopic velocity gradient. We find that the simulations
predicted values that are larger, by about a factor of two, than
our result for the Green-Kubo relation using experimental input.

The discrepancy between our result here and the simulation results
in Fig.~5 could arise from the differences between the simulations
and the experiment~\cite{Feng:10}. These differences include the
use of periodic boundary conditions in the simulations to mimic an
infinite system, while the dust particles in the experiment fill a
finite region due to dc radial electric fields. While all
particles in the simulations are identical, those in the
experiment vary in diameter by a few percent~\cite{Liu:03}, with a
similar variation in charge. Nonuniformities occur in the
simulations only as transients due to fluctuations, while the
experiment has static nonuniformities due to the dc radial
electric fields. These dc fields induce static stresses that can
lead to more structural disorder
which can result in an easier deformation of the arrangement of
dust particles and a reduction in the stress required to generate
shear flow, which is equivalent to a reduction of the viscosity.

Another possible explanation for the discrepancy with the
simulation results is that the potential in the
experiment~\cite{Feng:10} may not be exactly a Debye-H\"{u}ckel
potential, as assumed in Sec.~IV~A. One alternative, instead of
assuming a particular form for the potential, could be an
empirically determined potential of mean force, calculated from an
experimentally-measured pair correlation function, as has been
proposed theoretically~\cite{deAngelis:01, Hansen:86}. An
advantage of this approach is that all physical processes that
affect the potential would be included in the empirical
result~\cite{deAngelis:01}.


\section{VIII. Conclusions}

The Green-Kubo relation for viscosity has been tested using an
input of experimental data. The value for the viscosity determined
by the Green-Kubo relation with the input of data from an
experiment~\cite{Feng:10} was compared to the value from a
previous experiment using a hydrodynamical
method~\cite{Nosenko:04}. In both experiments the physical system
was a quasi-2D dusty plasma, and the conditions were similar,
aside from the absence of a macroscopic velocity gradient in the
experiment~\cite{Feng:10} for the Green-Kubo result. We found that
the results agree as well as can be expected. This agreement
serves as our test of the Green-Kubo relation for viscosity of a
dusty plasma.

Additionally, we compared our result for the viscosity determined
by the Green-Kubo relation with the predictions of MD
simulations~\cite{Donko:06, Feng:11, Liu:05}. The results were as
consistent as expected, given the differences between the
simulations and the experiment~\cite{Feng:10} that provided our
input data.

Further tests are needed for other Green-Kubo relations. Because
the Green-Kubo relations for the various transport coefficients
are all different, a test of the Green-Kubo relation for
viscosity, as we have presented here, does not also serve as a
test of another Green-Kubo relation. In addition, tests for other
physical systems, such as the 2D systems mentioned in Sec.~I,
would be useful.

This work was supported by NSF and NASA.

\begin{center}
\begin{table*}[b]
\caption{\label{tab:table1} Results for the kinematic viscosity
$\eta/\rho$, which is normalized here by $a^2\omega_{pd}$ to make
it dimensionless. Viscosity $\eta$ is reported as the mean for
multiple runs; the standard deviation $\sigma$ and standard
deviation of the mean $\sigma_M$ are listed.}
\begin{ruledtabular}
\begin{tabular}{ l c r r c r c c c}

data source & calculation & \multicolumn{4}{c}{data size} &
\multicolumn{3}{l} {$\eta/\rho$, units ($a^2\omega_{pd}$)}\\
\cline{3-6}\cline{7-9}
            & procedure   & $M$&$N$& run duration &runs& mean  & $\sigma$ & $\sigma_M$  \\ \hline

\\experiment~\cite{Feng:10}    & Eq.(5)$\rightarrow$(6)$\rightarrow$(7) & $\approx 600$    &$\approx 2100$ &$\approx 607\omega^{-1}_{pd} (20.2~{\rm sec})$ &4 &${\bf 0.16}$& ${\bf 0.04}$&${\bf 0.02}$
\\Langevin simulation (Sec.~V) & Eq.(5)$\rightarrow$(6)$\rightarrow$(7) & 600&4096 &$607\omega^{-1}_{pd}$&48&$0.24$&$0.08$&$0.01$
\\equilibrium simulation~\cite{Feng:11} & Eq.(1)$\rightarrow$(6)$\rightarrow$(7)& 4096&4096 &$1.86\times10^4\omega^{-1}_{pd}$&4&$0.26$&$0.02$&$0.01$
\\Langevin simulation~\cite{Feng:11}& Eq.(1)$\rightarrow$(6)$\rightarrow$(7) & 4096&4096 & $1.86\times10^4\omega^{-1}_{pd}$&4&$0.27$&$0.02$&$0.01$\\

\end{tabular}
\end{ruledtabular}
\end{table*}
\end{center}

\begin{figure}[p]
\caption{\label{setup} (color online) Configuration for the
experiment~\cite{Feng:10}. (a) This diagram of the vacuum chamber
is shown in an exploded view, to better show the lower electrode,
which in the experiment was located inside the chamber. A
low-pressure gas of neutral argon atoms filled the chamber. A
radiofrequency voltage was applied between two electrodes,
separated by an insulator. One electrode was the powered lower
electrode, and the other consisted of the grounded vacuum chamber
and shield. The gas was partially ionized, yielding a plasma with
three components: electrons, positive argon ions, and neutral
argon atoms. The $x$ and $y$ axes correspond to the two orthogonal
directions used in measuring the positions and velocities of dust
particles. The side ports were used to admit laser beams, not
shown here. For heating, a pair of 532~nm laser beams were
directed into the chamber by scanning mirrors as
in~\cite{Nosenko:06}, while for illumination a 577~nm laser sheet
was used with a configuration as in~\cite{Melzer:2000}. (b) This
sketch shows a side view of the lower electrode. Polymer
microspheres were introduced by shaking them into the plasma from
above, and they moved downward due to gravity $g$. They gained a
negative electric charge $Q$ and were levitated upward due to a
vertical dc electric field $E$ so that they remained in a single
horizontal layer above the lower electrode. There was also a
smaller radial dc electric field, not shown, which provided
horizontal confinement. Images of the dust particles were recorded
by a video camera viewing through the top port. Shown here are
$5~{\rm{mm}}\times5~{\rm{mm}}$ portions of two images: (c) a
Wigner lattice without laser heating (lines have been drawn to
indicate the lattice structure) and (d) a liquid maintained by
laser heating.}
\end{figure}


\begin{figure}[p]
\caption{\label{sketch} (color online). Sketch of the division of
the camera FOV into inner and outer regions for the
experiment~\cite{Feng:10}. In Eq.~(\ref{SSexp}), the subscripts
$i$ and $j$ refer to particles located in the inner region, and
both the inner and outer regions, respectively. The circle
indicates the cutoff distance $5 b$ for the potential. The unused
portion of the camera FOV on the right is not included in the
analysis.}
\end{figure}

\begin{figure}[p]
\caption{\label{SSexperiment} (color online). Stress
autocorrelation function SACF, i.e., $C_{\eta}(t)$, and its time
integration. Results are shown for four runs of the experiment
\cite{Feng:10}. All quantities shown are normalized. The SACF is
normalized as  ${\Bbb C} \equiv C_{\eta} / (Ak_BT\rho
a^2\omega_{pd}^2)$ and  is drawn here at $10\times$ magnification.
Time is given in units of $\omega_{pd}^{-1}$, that is, $\tau = t
\omega_{pd}$. The integral shown by the smooth curve is used in
the Green-Kubo relation, Eq.~(\ref{etaEXP}), to calculate the
viscosity. Choosing the integration limit $t_I$ as the time when
${\Bbb C}(\tau)$ crosses zero as described in Sec.~IV~D, the
integral yields the dimensionless viscosity, as indicated by the
solid circle for each run.}
\end{figure}

\begin{figure}[p]
\caption{\label{etameasurement} (color online). Our value of the
kinematic viscosity calculated from the Green-Kubo relation, using
input from the experiment~\cite{Feng:10}, compared to results from
a hydrodynamical analysis of a previous
experiment~\cite{Nosenko:04}. Values are made dimensionless by
normalizing by $a^2\omega_{pd}$. The $x$ axis, which has a
logarithmic scale, is the Coulomb coupling parameter $\Gamma$ as
defined in Sec.~III. Our result, shown as a solid diamond, is the
mean for four experimental runs in Fig.~3, and the vertical error
bar indicates only the run-to-run variation, calculated as the
standard deviation of the mean. The horizontal error bar (for the
result from the Green-Kubo relation) reflects a $10\%$ uncertainty
in $Q$.}
\end{figure}

\begin{figure}[p]
\caption{\label{etasim} (color online).  Comparison of our result
using the Green-Kubo relation for viscosity with the input of data
from the experiment \cite{Feng:10}, shown as a diamond as in
Fig.~4, to values from previously-reported 2D Debye-H\"{u}ckel
simulations~\cite{Donko:06, Feng:11, Liu:05}. The simulation
results shown for \cite{Feng:11} are also listed in Table~I. Both
axes are logarithmic.}
\end{figure}


\begin{thebibliography}{16}
\bibitem{error-test}
We carried out tests that showed that the SACF is more sensitive
to an error in $Q$ than in $\lambda_D$.
\bibitem{footnote-memory-time}
For comparison, the memory time of collective motion in this
liquid of dust particles was $\approx 10~\omega_{pd}^{-1}$, as
measured by the zero-crossing of the correlation function
$C_{\eta}(t)$.
\end{thebibliography}
\end{document}